\documentstyle[11pt,aaspp]{article}

\def\t0{\theta_{\circ}}

\def\be{\begin{equation}}
\def\en{\end{equation}}

\begin{document}

\title {Dust in the 55 Cancri planetary system}
\author{Ray Jayawardhana\altaffilmark{1,2,3},
Wayne S. Holland\altaffilmark{2,4},
Jane S. Greaves\altaffilmark{4},
William R. F. Dent\altaffilmark{5},\\
Geoffrey W. Marcy\altaffilmark{6},
Lee W. Hartmann\altaffilmark{1},
and Giovanni G. Fazio\altaffilmark{1}}
\altaffiltext{1}{Harvard-Smithsonian Center for Astrophysics, 60 Garden St., Cambridge, MA 02138; Electronic mail: rayjay@cfa.harvard.edu}
\altaffiltext{2} {Visiting Astronomer, James Clerk Maxwell Telescope,
which is operated by The Joint Astronomy Centre on behalf of the
Particle Physics and Astronomy Research Council of the United Kingdom, the
Netherlands Organisation for Scientific Research, and the National
Research Council of Canada.}
\altaffiltext{3} {Visiting Astronomer, W.M. Keck Observatory, which is
operated as a scientific partnership among the California Institute of
Technology, the University of California, and the National Aeronautics
and Space Administration. The Observatory was made possible by the
generous financial support of the W.M. Keck Foundation.}
\altaffiltext{4}{Joint Astronomy Centre, 660 N. A'ohoku Place, University
Park, Hilo, HI 96720}
\altaffiltext{5}{UK Astronomy Technology Centre, Royal Observatory, Blackford Hill, Edinburgh EH9 3HJ, United Kingdom}
\altaffiltext{6}{Department of Astronomy, University of California, Berkeley,
CA 94720}

\begin{abstract}
The presence of debris disks around $\sim$ 1-Gyr-old main sequence stars
suggests that an appreciable amount of dust may persist even in mature
planetary systems. Here we report the detection of dust emission from 55
Cancri, a star with one, or possibly two, planetary companions detected
through radial velocity measurements. Our observations at 850$\mu$m and
450$\mu$m imply a dust mass of 0.0008-0.005 Earth masses, somewhat higher
than that in the the Kuiper Belt of our solar system. The estimated
temperature of the dust grains and a simple model fit both indicate a central 
disk hole of at least 10 AU in radius. Thus, the region where the planets
are detected is likely to be significantly depleted of dust. Our results 
suggest that far-infrared and sub-millimeter observations are powerful tools 
for probing the outer regions of extrasolar planetary systems.
\end{abstract}


\section{Introduction}
Planetary systems are born in dusty circumstellar disks. Once planets
form, the circumstellar dust is thought to be continually replenished
by collisions and sublimation (and subsequent condensation) of larger 
bodies such as asteroids, comets, and Kuiper Belt objects (Nakano 1988; 
Backman \& Paresce 1993). Such debris disks have now been directly imaged
around several nearby main sequence stars: $\beta$ Pictoris, HR 4796A,
Vega, Fomalhaut and $\epsilon$ Eridani (Holland et al. 1998; Jayawardhana
et al. 1998; Koerner et al. 1998; Greaves et al. 1998). The presence of
debris disks around stars which, in some cases, may be $2\times10^8$ -
$10^9$ years old suggests that an appreciable amount -- perhaps tens of
lunar masses -- of dust may be present even in mature planetary systems.
The ring of dust recently imaged at 850$\mu$m around the nearby K2V star
$\epsilon$ Eridani is also spatially analogous to the Kuiper Belt of our
own solar system (Greaves et al. 1998). 
 
The G8V star 55 Cancri, at a distance of 13 pc, is unique in having both
planets as well as a substantial dust disk. It contains one planet of
about 2 Jupiter masses in an orbit with a semi-major axis of 0.11 AU
(Butler et al. 1997) and  evidence for a second planet at several AU in
the form of a residual drift in the stellar velocity over the past 10
years (Marcy \& Butler 1998). The dust disk, first inferred by Dominik et
al. (1998) using ISO observations at 25$\mu$m and 60$\mu$m, is much
larger, with a radius of $\sim$ 50 AU. Recent near-infrared coronographic
observations of Trilling \& Brown (1998) have resolved the 55 Cancri dust
disk and confirm that it extends to at least 40 AU (3.24'') from the
star.  

We have recently commenced a mini-survey of the parent stars of known
extrasolar planets using the Submillimeter Common User Bolometer Array
(SCUBA) on the James Clerk Maxwell Telescope (JCMT). Our program is to
obtain 850$\mu$m and 450$\mu$m flux measurements in the photometry mode
since the expected disk sizes are too small to be spatially resolved at
present. Our goals are to explore the kinship between circumstellar dust and
planets and to provide significant constraints on the nature and amount of
dust associated with the Kuiper Belts of these extrasolar planetary systems.
Here we report the detection of sub-millimeter emission from 55 Cnc during
the first observing shift our survey program. 

\section{Observations and Results}
We observed 55 Cnc with the SCUBA instrument (Holland et al. 1999) on the
JCMT on Mauna Kea, Hawaii. The data were obtained on 1999 February 4-9 UT
using the SCUBA photometry mode. Although SCUBA operates at 450 and
850$\mu$m simultaneously, the observing conditions are generally poorer
at the shorter wavelength. Zenith atmospheric opacities were exceptionally 
good at 850$\mu$m, ranging from 0.10 to 0.15. Observations of
Uranus were used for calibrations. Pointing accuracy was 2'', which is small
compared with the beam size of 15'' at 850$\mu$m (FWHM) and 8'' at
450$\mu$m. The data were reduced using the SCUBA User Reduction Facility
(Jenness \& Lightfoot 1998). 

We also obtained mid-infrared observations of 55 Cnc on 1999 May 3 UT 
using the OSCIR instrument on the Keck II telescope. OSCIR is a mid-infrared 
imager/spectrometer built at the University of Florida\footnote[1]{
Additional information on  OSCIR is available on the Internet at 
www.astro.ufl.edu/iag/.}, using a 128$\times$128 Si:As Blocked Impurity
Band (BIB) detector developed by Boeing. On Keck II, OSCIR has a plate
scale of 0.062''/pixel, providing a 7.9''$\times$7.9'' field of view. We 
used a chop frequency of 4 Hz and a throw of 8''. Images were obtained in 
N(10.8 $\mu$m) and IHW18(18.2 $\mu$m) filters, with on-source
integration times of 120 sec and 300 sec, respectively. The standard
stars $\mu$ UMa and $\alpha$ Boo were used for flux calibration.

In the sub-millimeter, we measure 2.8$\pm$0.5 mJy at 850$\mu$m
and 7.9$\pm$4.2 mJy at 450$\mu$m from 55 Cnc, presumably due to
thermal emission of dust in a Kuiper Belt-like population. In the 
mid-infrared, where the emission is dominated by the stellar photosphere,
we measure 1.0$\pm$0.1 Jy at 10.8$\mu$m and 280$\pm$28 mJy at 18.2$\mu$m.
The mid-infrared images do not show any evidence for spatial extension.
This is not surprising given that 55 Cnc has little or no excess above
the photosphere at these wavelengths. Table 1 lists all available
mid-infrared to sub-millimeter flux measurements and limits from our
observations, $IRAS$, and $ISO$.

\section{Discussion}
Following Backman \& Gillett (1987), we can write the fractional
luminosity of dust as $\tau = L_{d}/L_{*}$, where $L_{d}$ and $L_{*}$
are the luminosities of the dust debris and the star, respectively.
For 55 Cnc, based on its far-infrared excesses, 
$\tau \approx 7 \times 10^{-5}$, some two orders of magnitude lower than 
that of the debris disk prototype $\beta$ Pictoris. 

Figure 1 shows that a single-temperature blackbody can match the far-infrared 
and sub-millimeter flux measurements of 55 Cnc quite well. If one assumes that 
the emission at $\lambda \leq$ 25$\mu$m is primarily due to the stellar 
photosphere, a $T=100$ K blackbody fits the ISO 60$\mu$m measurement and 
the SCUBA 450 and 850$\mu$m detections. It is also roughly consistent with 
the ISO 90$\mu$m limit. Figure 1 also includes a modified blackbody fit with 
$T=60$ K and $\beta =$0.5, where $F_{\nu} \propto \nu^{2+\beta}$, for 
comparison. This can fit the 60$\mu$m and sub-millimeter points well, but 
does not meet the ISO 90$\mu$m constraint.

A blackbody temperature of 100 K for the grains suggests that the
dust disk around 55 Cnc has a central disk hole with a minimum radius of
13 AU. If the dust temperature is closer to 60 K, the hole could be
as large as 35 AU. This would not conflict significantly with the
coronograph image of Trilling \& Brown (1998) where reliably detected
emission begins at $\sim$ 27 AU from the star. Thus, the 55 Cnc dust disk is
likely to be well outside the orbits of the two known planets in the system. 
Since $\beta$=0 corresponds to large grains and $\beta$=1 to small grains for 
optically thin emission, our best-fit to the data in Figure 1 would imply a 
population of grains with $a \sim 100 \mu m$.  

To better constrain the disk parameters using data at all wavelengths,
we used the model discussed
by Dent et al. (1999). Calculations are made with a 2-D continuum radiative
transfer code which includes the star and a thin disk with inner and outer
boundaries $r_{in}, r_{out}$, and a power-law mid-plane density $r^{-p}$.
The dust emission is characterized by a single characteristic grain size
$a$, a critical wavelength $\lambda_0$ and an opacity index $\beta$;
shortwards of $\lambda_0$ the grains act as blackbodies while longwards the
emissivity is given by $Q \propto (\lambda/\lambda_0)^{-\beta}$.

We have assumed the $r^{-3}$ power-law density distribution derived from
the near-infrared observations also continues down to the inner radius 
$r_{in}$. For 55 Cnc, the best fit model (Figure~2) has $r_{in}$ of 10 AU, 
a grain size $a$ = 100 $\mu$m and opacity index $\beta = 0.5$. 
Both the simple fit and model are roughly consistent with
the ISO 90 $\mu$m upper limit, although the lower $\beta$ may provide
a better fit to this data. The model indicates an upper limit to
the dust density in the region 3\,AU$\leq r_{in} \leq$10\,AU
of $<$10\% of the density at 10\,AU, thus
the region where planets are detected is significantly depleted of dust.

Since the sub-millimeter flux is less sensitive to the temperature of the 
grains than the infrared flux, we can use it to estimate the dust mass.
Following Jura et al. (1995), the dust mass $M_d$ is given by

\begin{equation}
M_d = F_{\nu} R^2 \lambda^2 / \left [2kT_{gr} K_{abs} \left (\lambda \right )\right ],
\end{equation}
if $R$ denotes the distance from the sun to 55 Cnc.  Assuming a dust 
absorption coefficient $K_{abs}(\lambda)$ between 1.7 and 0.4 cm$^2$ g$^{-1}$
at 850$\mu$m (Greaves et al. 1998), we obtain a dust mass of 0.0008-0.005 Earth
masses, for $T$= 100-130 K. The lower value of $K_{abs}(\lambda)$ is suggested
by models of large, icy grains (Pollack et al. 1994), while the higher estimate
has been used for previous observations of debris disks (Holland et al. 1998).
However, as for all the extrasolar debris disks, very large grains could
dominate the total mass while adding little submillimeter emission, so our
mass estimates only provide lower limits.  

Our dust mass estimates are consistent with Dominik et al. (1998) who derive 
$M_d > 4 \times 10^{-5} M_{Earth}$ by fitting a disk model to the ISO and 
IRAS data. On the other hand, using a low albedo (near-infrared reflectance 
of 6\%) and an average particle density of 1 g cm$^{-3}$, Trilling \& Brown 
(1998) estimate a dust mass of 0.4 $M_{Earth}$ in the 55 Cnc disk from their 
scattered light observations. Their estimate is inconsistent with ours.
The reason for the discrepancy is not clear. One possibility is that Trilling 
\& Brown (1998) may have overestimated the disk brightness in the 
near-infrared due to difficulties in background subtraction. 

The amount of dust in our solar system's Kuiper Belt is not well determined.
Based on far-infrared observations of $COBE$ and $IRAS$, Backman, Dasgupta
\& Stencel (1995) and Stern (1996) have placed an upper limit of 10$^{-5}
M_{Earth}$ on the Kuiper Belt mass in the form of dust (particles with
$a \le$ 1cm). However, Teplitz et al. (1999) show that the dust mass could
be as high as $7 \times 10^{-4} M_{Earth}$ depending on assumptions about
albedo, distribution in particle size, contribution of foreground and 
background sources to the far-infrared emission, etc. Thus, the
55 Cnc disk may be somewhat ``dustier'' than our Kuiper Belt. It also 
appears somewhat ``over-dusty'' for 55 Cnc's stellar age of $\sim$ 5 Gyr 
(Gonzalez \& Vanture 1998; Baliunas et al. 1997), when compared
to dust masses in the handful of nearby, well-studied debris disks.

Trilling \& Brown (1998) suggest that the apparent dust mass excess in
55 Cnc is consistent with the idea that the inner planet migrated toward
the star from its birthplace (Trilling et al. 1998; Murray et al. 1998). 
In this scenario, a planet migrates inward by exchanging angular momentum
with a circumstellar disk which initially extends to a few stellar radii. 
This migration could transfer material from the inner part of the disk to
the outer part, enhancing the mass at Kuiper Belt distances. If that is
true, other extrasolar planetary systems with ``hot Jupiter'' planets
should also harbor appreciable amounts of dust in their outer regions. We
expect to test this prediction during our on-going sub-millimeter survey
of parent stars of radial-velocity planets.

The radiation field of a G8 star is generally too weak to expel dust grains
by radiation pressure. In the case of 55 Cnc, the Poynting-Robertson timescale 
is much shorter than the estimated $\sim$5-Gyr age of the star (Dominik et
al. 1998). Therefore, the dust grains in the system must be replenished
by collisions or sublimation of larger bodies such as asteroids, comets, 
and Kuiper Belt objects. 

In summary, we have detected sub-millimeter thermal emission from dust 
in the 55 Cnc planetary system. Our results confirm that state-of-the-art 
sub-millimeter instruments are able to detect continuum emission from even 
a relatively small amount of dust surrounding nearby sun-like stars. The 
observed dust mass in the 55 Cnc system appears to be somewhat higher than 
that associated with the Kuiper Belt in our solar system. Far-infrared
observations from the {\it Space Infrared Telescope Facility} and the
{\it Stratospheric Observatory for Infrared Astronomy} as well as detailed
modeling will be crucial for reliably constraining the spatial extent, 
size distribution and composition of the dust. 

\bigskip
We are grateful to Charles Telesco, Scott Fisher and 
Robert Pi\~na for their assistance with the OSCIR observations at Keck. 
We also wish to thank the staff of JCMT and Keck for their outstanding
support. 

\newpage
\begin{table}
\begin{scriptsize}
\begin{center}
\renewcommand{\arraystretch}{1.2}
\begin{tabular}{lcc}
\multicolumn{3}{c}{\scriptsize TABLE 1}\\
\multicolumn{3}{c}{\scriptsize MEASURED FLUX DENSITIES FOR 55 CNC}\\
\hline
\hline
$\lambda (\mu m)$ & F$_{\nu}$ (mJy) & Source\\
\hline
\hline
10.8 & 1000$\pm$100 & OSCIR/Keck\\
12   & 981	    & IRAS\\	
18.2 & 280$\pm$28   & OSCIR/Keck\\
24   & 170$\pm$20   & ISO\\	
25   & 241  	    & IRAS\\
60   & $<$400       & IRAS\\
61   & 170$\pm$30   & ISO\\
95   & $<$90	    & ISO\\	
100  & $<$1000      & IRAS\\
161  & $<$780	    & ISO\\
185  & $<$750       & ISO\\
450  & 7.9$\pm$4.2  & SCUBA/JCMT\\
850  &  2.8$\pm$0.5 & SCUBA/JCMT\\
\hline
\end{tabular}
\end{center}
\end{scriptsize}
\end{table}

\newpage

\centerline{\bf Figure Captions}

\bigskip
\bigskip

Figure 1. Composite spectral energy distribution (SED) of 55 Cancri  
from infrared to sub-millimeter wavelengths. The near-infrared fluxes (open 
circles) are from Persson et al. (1977). ISO measurements are shown as
filled circles and ISO upper limits as open circles, while filled squares
and open squares designate IRAS measurements and upper limits, respectively.
Our mid-infrared fluxes are shown as diamonds and our JCMT measurements are 
indicated by filled stars. All error bars are smaller than the symbols except 
where shown. Also shown are the photospheric emission with T$_*$=5250\,K 
(solid line), and modified black bodies with $\beta$=0.0, T=100\,K (dashed 
line), and $\beta$=0.5, T=60\,K (dotted line), constrained to fit the 
850$\micron$ flux.

Figure 2. Best-fit model for the 55 Cnc SED. The model assumes a thin disk
with an inner radius $r_{in}=$10 AU, a grain size $a=$100 $\mu$m, and an 
opacity index $\beta=$0.5. Symbols are the same as in Figure 1.

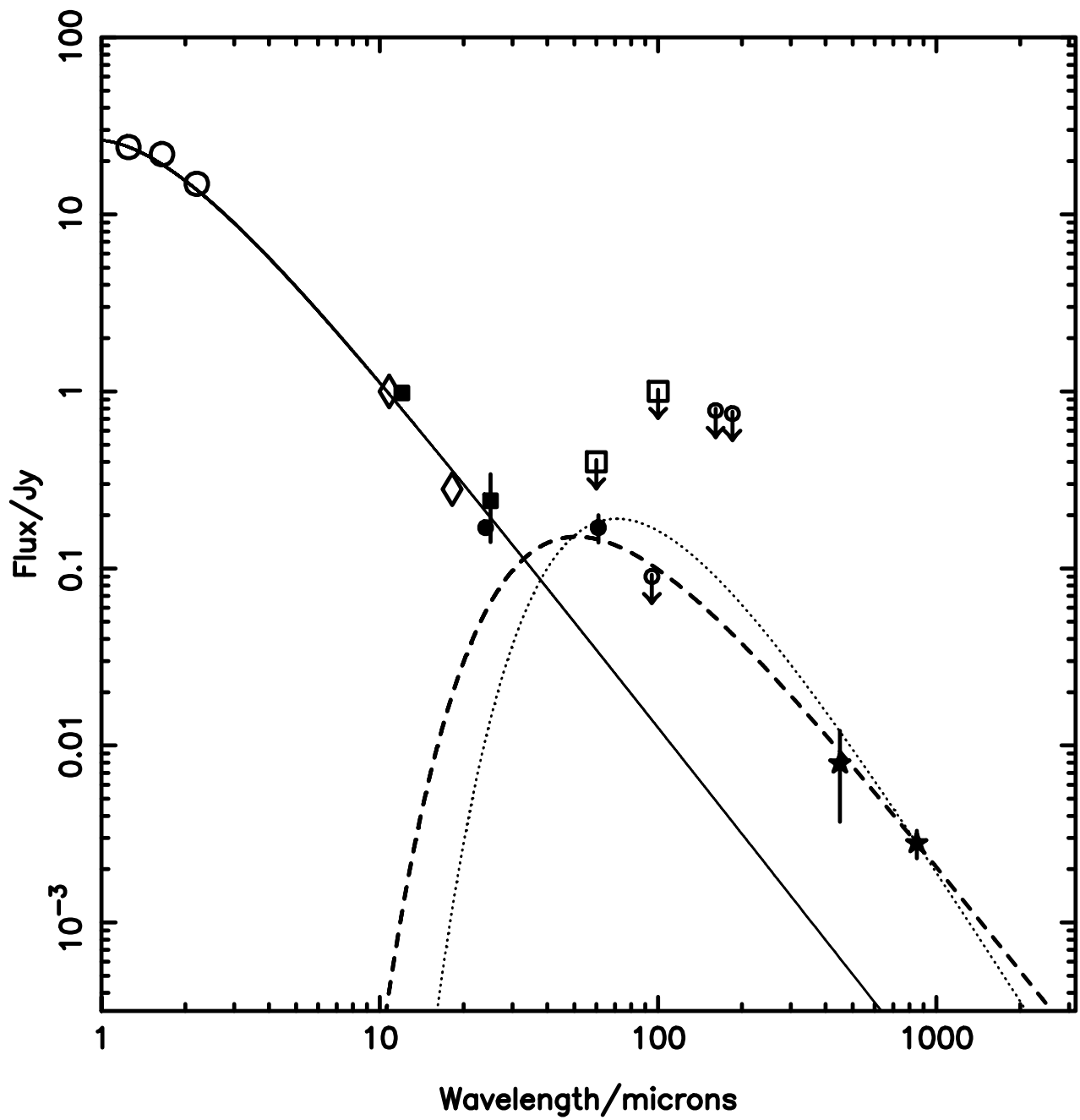
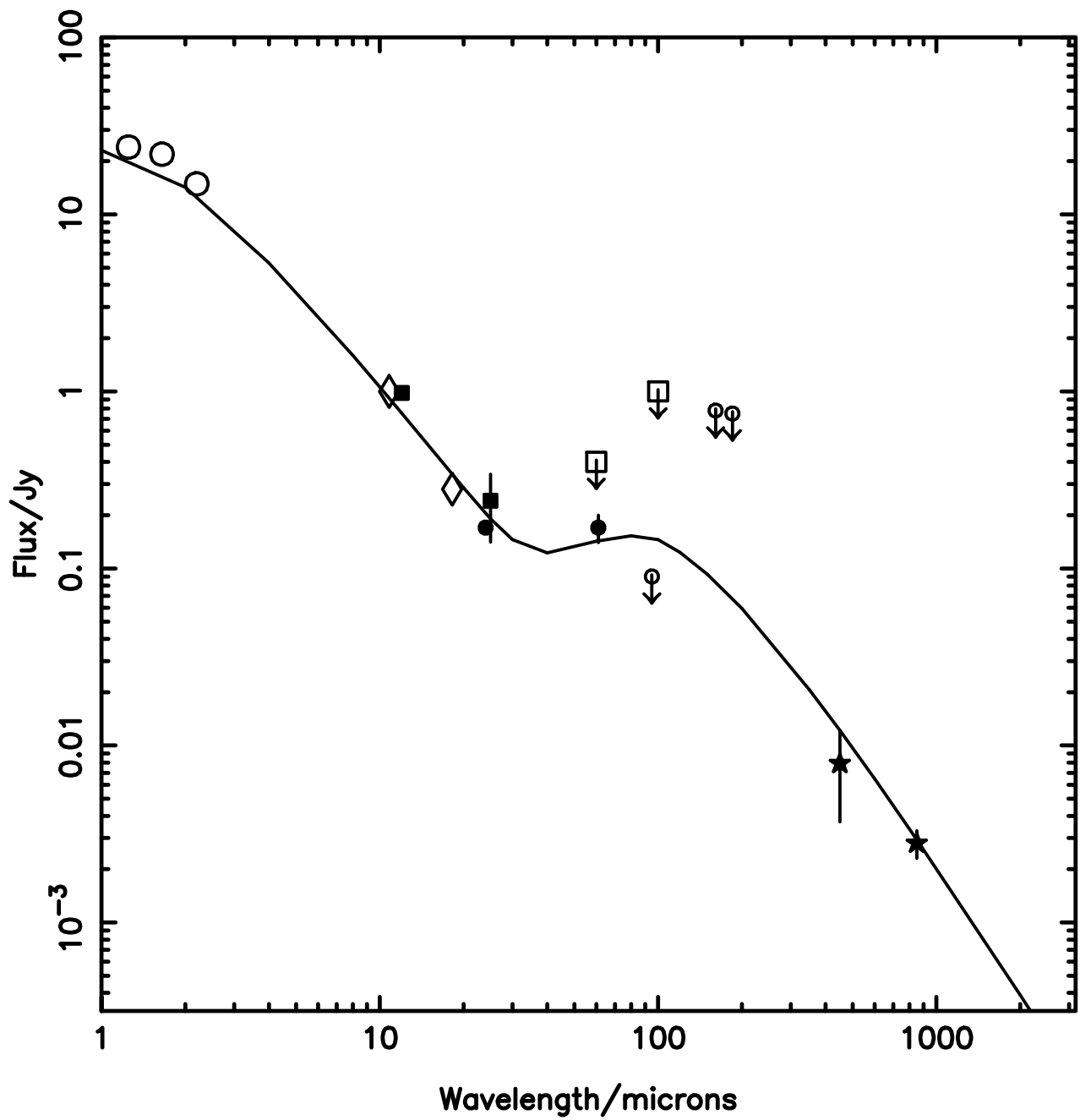
\end{document}